\documentclass[10pt]{wlscirep2}
\usepackage{tikz}
\usetikzlibrary{arrows,calc}
\definecolor{matlabBlue}{rgb}{ 0.0000, 0.4470, 0.7410}
\definecolor{matlabRed}{rgb}{ 0.8500, 0.3250, 0.0980}
\definecolor{matlabYellow}{rgb}{ 0.9290, 0.6940, 0.1250}
\usepackage{subcaption}
\newcommand{\om}{\omega'}
\title{All-depth dispersion cancellation  in spectral domain optical coherence tomography using numerical intensity correlations}

\author[1,*]{Mikkel Jensen}
\author[1]{Niels Møller Israelsen}
\author[2,3]{Michael Maria}
\author[2]{Thomas Feuchter}
\author[3]{Adrian Podoleanu}
\author[1,2]{Ole Bang}
\affil[1]{Technical University of Denmark, DTU Fotonik, Kongens Lyngby, 2800, Denmark}
\affil[2]{NKT Photonics, Birkerød, 3460, Denmark}
\affil[3]{University of Kent, School of Physical Sciences, Canterbury, Kent, England, CT2 7NZ}

\affil[*]{mikkje@fotonik.dtu.dk}

\begin{abstract}
In ultra-high resolution (UHR-) optical coherence tomography (OCT) group velocity dispersion (GVD) must be corrected for in order to approach the theoretical resolution limit. One approach promises not only compensation, but complete annihilation of even order dispersion effects, and that at all sample depths. This approach has hitherto been demonstrated with an experimentally demanding ’balanced detection’ configuration based on using two detectors.
We demonstrate intensity correlation (IC) OCT using a conventional spectral domain (SD) UHR-OCT system with a single
detector. IC-SD-OCT configurations exhibit cross term ghost images and a reduced axial range, half of that of conventional
SD-OCT. We demonstrate that both shortcomings can be removed by applying a novel generic artefact reduction algorithm and using analytic interferograms. We show the superiority of IC-SD-OCT compared to conventional SD-OCT by showing how
IC-SD-OCT is able to image spatial structures behind a strongly dispersive silicon wafer.
Finally, we question the resolution enhancement of $\sqrt{2}$ that IC-SD-OCT is often believed to have compared to SD-OCT. We
show that this is simply the effect of squaring the reflectivity profile as a natural result of processing the product of two intensity spectra instead of a single spectrum.
\end{abstract}
\begin{document}

\flushbottom
\maketitle
\thispagestyle{empty}

\section*{Introduction}
In-depth imaging of human tissue has been one of the greatest achievements of optical technologies. Optical coherence tomography (OCT) was initiated more than 25 years ago, when a cross-sectional image of the human retina using a Michaelson interferometer was demonstrated\cite{Huang1991}. The ability to display changes of the refractive index by detecting photons balistically backscattered millimetres inside tissue at the micrometer scale has let to a revolution in the field of ophthalmology and is essential for many other medical fields \cite{OCT2015}. Quite recently OCT has even been demonstrated for macroscopic imaging, thereby adding a new perspective in terms of its applications \cite{wang2016cubic}.\\ 
With a Gaussian spectral profile the axial (in-depth) resolution limit is intrinsically given by $\delta z = \frac{2\ln(2)}{\pi}\frac{\lambda_c^2}{\Delta\lambda}$, where $\lambda_c$ is the central wavelength and $\Delta\lambda$ is the full-width at half maximum spectral bandwidth (FWHM) of the light source \cite{OCT2015}. To maximize the penetration depth in tissue $l_c$ is typically chosen to be in the near infrared (NIR) regime to minimize scattering \cite{OCT2015}, but well below the major absorption bands of water peaking at $\lambda=3$~$\mu$m\cite{curcio1951near}. In order to maintain $\delta z$ when increasing the wavelength from the visible to the NIR, one is left to maximize $\Delta \lambda$. In doing so, chromatic dispersion in both optical components and sample will degrade the depth resolution. This is due to each wavelength experiencing a different optical path through the system and sample, causing the optical path difference to differ as well. The effect of the different optical path lengths in the two paths (reference arm and sample arm) for different wavelengths is commonly known as the dispersion mismatch \cite{hitzenberger1999dispersion,OCT2015}.\\ \indent
To counter the dispersion mismatch, dispersion compensation (DC) is done hardware wise by ensuring that the two arms are constructed identically. However, this makes the set-up more costly and increases complexity. Instead simple DC with a glass plate, such as BK7, is today used to balance the dispersion \cite{hitzenberger1999dispersion}. Alternatively, a large variety of numerical approaches have been introduced, first for time domain (TD) OCT \cite{fercher2001numerical,fercher2002dispersion,marks2003digital} and later for spectral domain OCT (SD-OCT). In particular, in SD-OCT several new DC methods have been demonstrated \cite{wojtkowski2004ultrahigh,cense2004ultrahigh,makita2008full,hillmann2012common,choi2012extracting,Bradu:15}, that can achieve single-interface DC,  i.e., sharpening only one interface in the sample. Only a few methods promise multi-interface DC, which is necessary in order to maintain axial resolution throughout the imaging depth of a multi-layered dispersive sample \cite{lippok2012dispersion, Pan:17,nasr2003demonstration,nasr2004dispersion,erkmen2006phase,le2010experimental,kaltenbaek2008quantum,lavoie2009quantum}. One approach that can compensate only second order dispersion at multiple interfaces, is the fractional Fourier transform combined with numerical segmentation of the sample and a radon transform, posing a heavy computational load, which scales with the number of pre-defined sample segmentations in depth \cite{lippok2012dispersion}. Another simpler approach is to perform a linear interpolation of the depth-dependent DC from two depths where the dispersion mismatch is known \cite{Pan:17}.\\
An alternative approach, inspired by quantum OCT\cite{nasr2003demonstration,nasr2004dispersion}, is  phase conjugate OCT\cite{erkmen2006phase,le2010experimental} and chirped-pulse interferometry OCT\cite{kaltenbaek2008quantum,lavoie2009quantum}. These approaches can do even-order dispersion cancellation, but are costly and complex hardware-wise, due to the requirement of sum frequency generation, while providing only low sensitivity.A numerical scheme exploiting a generalized auto-convolution function for depth-dependent dispersion cancellation, also developed in the foot steps of quantum OCT, was proposed by Banaszek et al. and termed 'blind dispersion compensation' \cite{banaszek2007blind}. This method promises protection from GVD using a conventional OCT system by numerically convolving two A-scans. Hardware implementations of Banaszek's approach have also been proposed and termed 'spectral intensity' or 'intensity' OCT \cite{resch2007classical,lajunen2009resolution,ryczkowski2016experimental,shirai2014intensity}, but these again require two detectors and added complexity of the experimental set-up. We here consider the numerical technique of Banaszek and apply it in the spectral domain using a single spectrometer. We term it intensity correlation spectral domain OCT (IC-SD-OCT).\\

All reports so far on implementing IC-SD-OCT share two major drawbacks compared to conventional SD-OCT: (1) Halving of the imaging depth, and (2) the appearance of IC artefacts stemming from intensity cross terms. Extending the numerical scheme of Banaszek et al, we here for the first time demonstrate single-detector ultra-high resolution SD-OCT with all-depth multi-interface sample dispersion removal with significant artefact reduction and full imaging depth. We do this by numerically implementing the IC scheme, but on the analytical signal, which we distinguish from the standard IC scheme by denoting it ICA. The novel artefact reduction scheme is then applied to the ICA signal. By imaging two different silicon phantoms, we highlight the applicability of IC-SD-OCT with a conventional SD-OCT set-up, and show that GVD is intrinsically removed at all depths of the sample with no depth segmentation or conventional DC needed, while maintaining the imaging depth. 
%
%
\\
\\
\section*{Theory}
\label{sec:theory}
In this section the theory behind IC-SD-OCT is presented. First, we introduce the basic concept of IC-SD-OCT in the settings of conventional OCT, and we later apply the analytic signal to explain the image depth-maintaining procedure. Subsequently, the full mathematical framework of ICA-SD-OCT is presented, including the artefact reduction technique. Finally we discuss the axial resolution in IC-SD-OCT, and show numerical simulations to validate the theoretical predictions, and ICA-SD-OCT is compared to quantum OCT.
\subsection*{Intensity correlation spectral domain optical coherence tomography - IC-SD-OCT}
In SD-OCT, the channelled spectrum (interferogram) is given by
\begin{equation}\label{eq:I}
I(\omega) \propto |E_R(\omega) + E_S(\omega)|^2 = |E_R|^2 + |E_S|^2 + E_RE_S^* + E_R^*E_S,
\end{equation}
where $E_R, E_S$  are the electric fields returned to the spectrometer from the reference arm and sample arm, respectively, see Fig. \ref{fig:ansig}(a) for a sketch of the set-up. The electric field from the reference arm is:
\begin{align}\label{eq:ER2}
E_R(\omega) &= \sqrt{\frac{I_0(\omega)}{2}}\exp\left( i\omega t - i k l_R \right),
\end{align}
and for two scattering centres in the sample arm, the electrical field from the sample can be written as:
\begin{align}
E_S(\omega) &= \sqrt{\frac{I_0(\omega)}{2}}\exp\left( i\omega t - i k l_S\right)\left[r_1\exp\left(- i \beta(\omega)L_1\right) + r_2\exp\left( -i \beta(\omega)L_2 \right)\right],
\label{eq:ES2}
\end{align}
where $I_0$ is the source spectrum, $r_1,r_2$ are the complex reflection coefficients, $l_S, l_R$ are the sample and reference paths, measured as twice the distance from the beam splitter to the sample surface and reference mirror, respectively. $L_1, L_2$ are twice the distances from the sample's surface to each of the scattering centres, and $\beta(\omega) = \frac{\omega}{c}n(\omega)$ is the wavenumber of the sample, with $c$ being the vacuum speed of light, and $n$ being the depth-averaged refractive index of the medium. In general the depth-averaged refractive index will of course be different for two scattering centres at different depths, but we assume this difference to be negligible, such that $n_1(\omega)\approx n_2(\omega)\equiv n(\omega)$. 
%
%
%
%
Assuming real reflection coefficients, equations \eqref{eq:ER2}, \eqref{eq:ES2} and \eqref{eq:I} are combined and the normalised interferogram, $I_n$ is obtained through $I_n = (I(\omega)-I_0(\omega)/2)/{(I_0(\omega)/2)}$, yielding:
\begin{equation}\label{eq:convOCT}
I_n(\omega) =\frac{2I(\omega)-I_0(\omega)}{I_0(\omega)}= r_1^2 + r_2^2 + 2r_1r_2\cos\left(\Delta L\beta(\omega)\right) + 2r_1\cos\left([\omega\Delta l/c + \beta(\omega)L_1]\right) + 2r_2\cos\left([\omega\Delta l/c + \beta(\omega)L_2]\right) ,
\end{equation}
where $\Delta L = L_2 - L_1$, and $\Delta l = l_S - l_R$. To generate the IC-SD-OCT interferogram, $I_n$ is multiplied by itself, however flipped around the central frequency, $\omega_0$, and complex conjugated:
\begin{equation}\label{eq:IC1}
I_{IC}(\omega_0,\omega') = I_n(\omega_0+\omega'){I_n}^*(\omega_0-\omega')
\end{equation}
where $\omega'=\omega-\omega_0$. This intra-spectral product between the two optical frequency components $\omega_0+\omega'$ and $\omega_0-\omega'$ can be understood as probing the sample and the reference object at two different frequencies and seeking cross correlations between all the four electric fields involved, hence fourth-order field correlations. This classical approach is inspired by quantum OCT directly measuring fourth order correlations, which we will discuss in a later section. It is important to note that as IC-SD-OCT is a classical analogy to quantum OCT, the first realizations supposedly required two spectrometers to mimic the two photo detectors of quantum OCT, but Shirai showed that it is fundamentally equivalent to using two identical spectra obtained by one spectrometer instead of two different spectrometers in a ’balanced detection’ configuration\cite{shirai2016improving}. This means that $I_n(\omega_0 - \omega')$ in Eq. \eqref{eq:IC1} can be obtained either experimentally or numerically by mirroring $I_n(\omega_0 + \omega')$.

Equation \eqref{eq:IC1} contains multiplication of cosines from equation \eqref{eq:convOCT}, which create oscillations with half the initial period. All peaks of the Fourier transform of equation \eqref{eq:IC1} are shifted to twice the OPD due to the decreased period of the oscillations. The spacing between points of discrete sampling of the spectrum in k-space, $\Delta k$, is fixed by the spectrometer, which fixes the full depth range (both positive and negative OPD) to $z_S=2\pi/\Delta k$. It is therefore possible that peaks that were positioned well below the Nyquist limit, $z_N=z_S/2$ before the multiplication are above after, reaching up to twice the Nyquist limit, as illustrated in Fig. \ref{fig:ansig}(b). This reduces the available depth range without aliasing (the Nyquist Sampling Theorem) by a factor of two \cite{ryczkowski2016experimental}. In addition, the cross terms from multiplication between different cosines cause artefacts, which deteriorate the image quality \cite{banaszek2007blind,shirai2014intensity,shirai2016improving,ryczkowski2016experimental}. For IC-SD-OCT to be relevant, the imaging depth must be restored, and the artefacts eliminated.

\subsection*{Restoring the imaging depth by use of the analytic signal - ICA-SD-OCT}
\begin{figure}
\centering
\includegraphics[width = \textwidth]{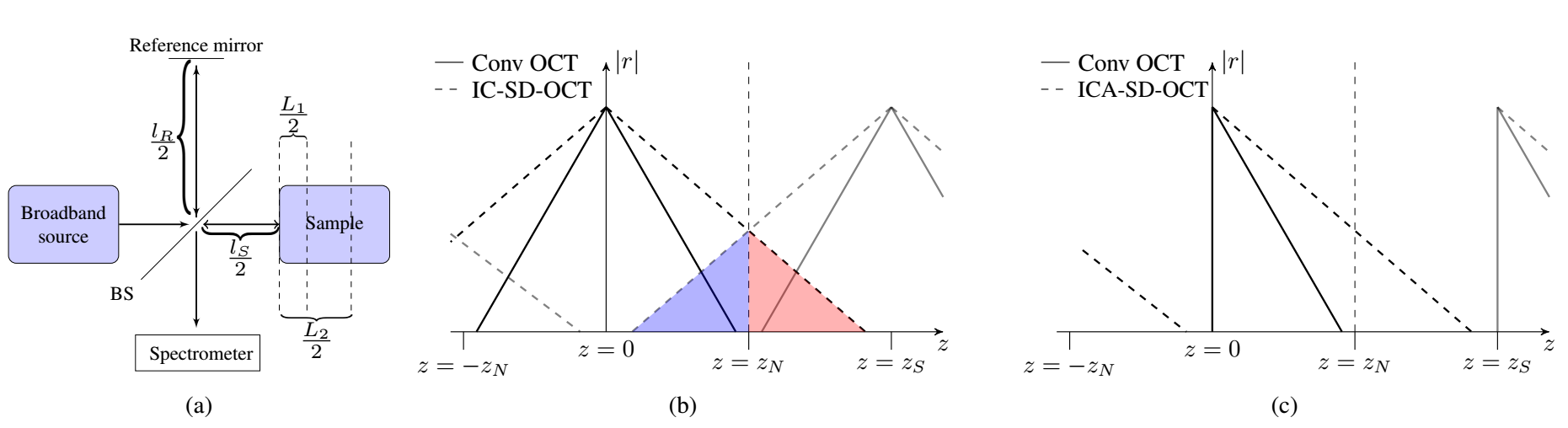}
\caption{(a) Schematic drawing of the Michaelson interferometer with the marked path length emplaoyed in the text. BS is a beam splitter. (b) Schematic illustration of an A-scan before and after the IC-OCT operation. In (a) the IC-OCT procedure is applied in a straight forward manner, and in (c), the IC-OCT procedure is applied to the analytic signal of the interference spectrum. The grey lines with a peak at $z = z_S$ are artefacts from the discrete sampling of the spectrum, and the observed A-scan is a sum of all parts, both artefacts and non-artefacts. This means that when the artefacts overlap on the real A-scan, aliasing occurs, which causes loss of information from the deep layers (red in (c), as well as obscuration of the otherwise still visible part of the A-scan (blue in (b)). In (c) the artefact does not interfere up to $z=z_S$, and the full range of points is used such that the imaging depth is maintained through the IC-OCT procedure.}
\label{fig:ansig}
\end{figure}

The coloured hatched areas in Fig. \ref{fig:ansig}(b) indicate the IC-SD-OCT signal and aliased signal trespassing into one another’s imaging range set by $z_N$ (dashed lines). To eliminate this aliasing problem we here, for the first time to our knowledge, propose to use the complex analytic interferograms in Equation (5), instead of the real-valued interferograms. The complex analytic signal $I_a$  of a real signal $I$ is computed by applying the Hilbert transform (HT), $\mathcal{H}\{f(\omega)\} \equiv 1/\pi\omega \: \otimes f(\omega)$,  

\begin{equation}
I_a(\omega) = I(\omega) + i\mathcal{H}\lbrace I(\omega) \rbrace,
\end{equation}
where $\times$ denotes convolution. The analytic signal is zero for negative OPDs by definition, and thus also for depths between $z_N$ and $z_S$ due to the repetition of the spectrum of discretely sampled signals, as illustrated in figure \ref{fig:ansig}(c) (solid line). The components of the IC-SD-OCT interferogram that are deeper than the Nyquist depth (red part in figure \ref{fig:ansig}(b)) are, when using the analytic signal in what we term the ICA scheme, will therefore be fully distinguishable, i.e., aliasing will be eliminated, as seen in Fig. \ref{fig:ansig}(c). As a result of any of the IC and ICA procedures, the density of points is doubled, but by using the ICA scheme, the imaging depth is maintained because all points are utilised and not only half.

Shirai has theoretically investigated the application of IC-SD-OCT for multiple scattering samples in the special case where dispersion originates from only a dispersive element in the sample arm of the SD-OCT system\cite{shirai2016improving}, i.e., neglecting the dispersion from the sample itself. Here we present an extended derivation that takes the dispersion from the sample into account. We want to derive a theory for multiple scatterers because the IC- and ICA-SD-OCT procedures in this case will create artefacts due to the multiplication in Eq. \eqref{eq:IC1} creating cross terms. We therefore consider the simplest case with cross terms, which is with two scatterers, without loss of generality. Using the analytic signal of equation \eqref{eq:convOCT}, Taylor expanding $\beta(\omega) \approx \beta_0 + \beta_1\om + \beta_2\om^2/2$, and defining $\beta' \equiv \beta_0 + \beta_2\om^2/2 + ...$ to include all even order dispersion terms, we find
\begin{equation}\label{eq:Ia}
I_{n,a}(\omega_0 + \omega') = r_1^2 + r_2^2 + 2r_1r_2\mathrm{e}^{i\Delta L[\beta' + \beta_1\omega']} + 2r_1\mathrm{e}^{i[(\omega_0 + \om)\Delta l/c + \beta'L_1 + \beta_1\omega'L_1]} + 2r_2\mathrm{e}^{i[(\omega_0 + \om)\Delta l/c + \beta'L_2 + \beta_1\omega'L_2]}.
\end{equation}
Neglecting higher order odd dispersion terms, $\beta_3, \beta_5, ...$, the ICA-SD-OCT spectrum is then obtained from Eq. \eqref{eq:IC1}
\begin{equation}\label{eq:ISI}
\begin{split}
I_{IC}(\omega',\omega_0) &= I_{n,a}(\omega_0 + \omega')I_{n,a}^*(\omega_0-\omega')\\
&=(r_1^2 + r_2^2)^2 + 4r_1^2r_2^2\mathrm{e}^{i2\om(\beta_1\Delta L +  \Delta l /c) } + 4r_1^2\mathrm{e}^{i2\om(\beta_1 L_1 + \Delta l /c)} + 4r_2^2\mathrm{e}^{i2\om(\beta_1 L_2+ \Delta l /c)}\\
&\hspace{5mm} + 4r_1r_2(r_1^2 + r_2^2)\mathrm{e}^{i\om(\beta_1 \Delta L + \Delta l/c)}\cos(\beta'\Delta L) + 4r_1^2(r_1+r_2)\mathrm{e}^{i\om(\beta_1 L_1 + \Delta l/c)}\cos(\omega_0\Delta l/c + \beta'L_1)\\
& \hspace{5mm} + 4r_2(3r_1^2 + r_2^2)\mathrm{e}^{i\om(\beta_1 L_2 + \Delta l/c)}\cos(\omega_0 \Delta l/c + \beta'L_2) + 8r_1r_2^2\mathrm{e}^{i\om(\beta_1[L_2 + \Delta L] + \Delta l/c)}\cos(\omega_0 \Delta l/c + \beta'[L_2 - \Delta L])\\
& \hspace{5mm}+ 8r_1r_2\mathrm{e}^{i\om(\beta_1[L_1+L_2] + \Delta l/c)}\cos(\beta' \Delta L),
\end{split}
\end{equation}
with * denoting complex conjugates. The four terms of the second line in equation \eqref{eq:ISI} are equivalent to the four terms from conventional OCT in equation \eqref{eq:Ia}, only positioned at twice the OPD and without any GVD from the dominant dispersion term $\beta_2$ and all other even orders of dispersion. In order to maintain the correct physical distances, the $z$ axis must be scaled by a factor $\frac{1}{2}$ as previously explained, and the point density is thus also increased by a factor of two.

\subsection*{Artefact reduction}
As discussed above $I_n(\omega_0-\omega')$ in Eq. \eqref{eq:IC1} can be obtained either experimentally or numerically from $I_n(\omega_0+\omega’)$.  The 'balanced detection' experimental configuration has been shown to suppress some of the artefacts in IC-SD-OCT\cite{ryczkowski2016experimental,shirai2014intensity}, i.e., some of the last 5 terms in Eq. \eqref{eq:ISI}. It has also been shown that in the numerical configuration that artefacts can also be removed, but only one at a time using a window function \cite{ryczkowski2016experimental,shirai2016improving}, which allows us to remove all artefacts numerically, thereby leaving the 'experimental balanced detection' redundant.

\begin{figure}
\centering
\includegraphics[height = 0.5\textwidth]{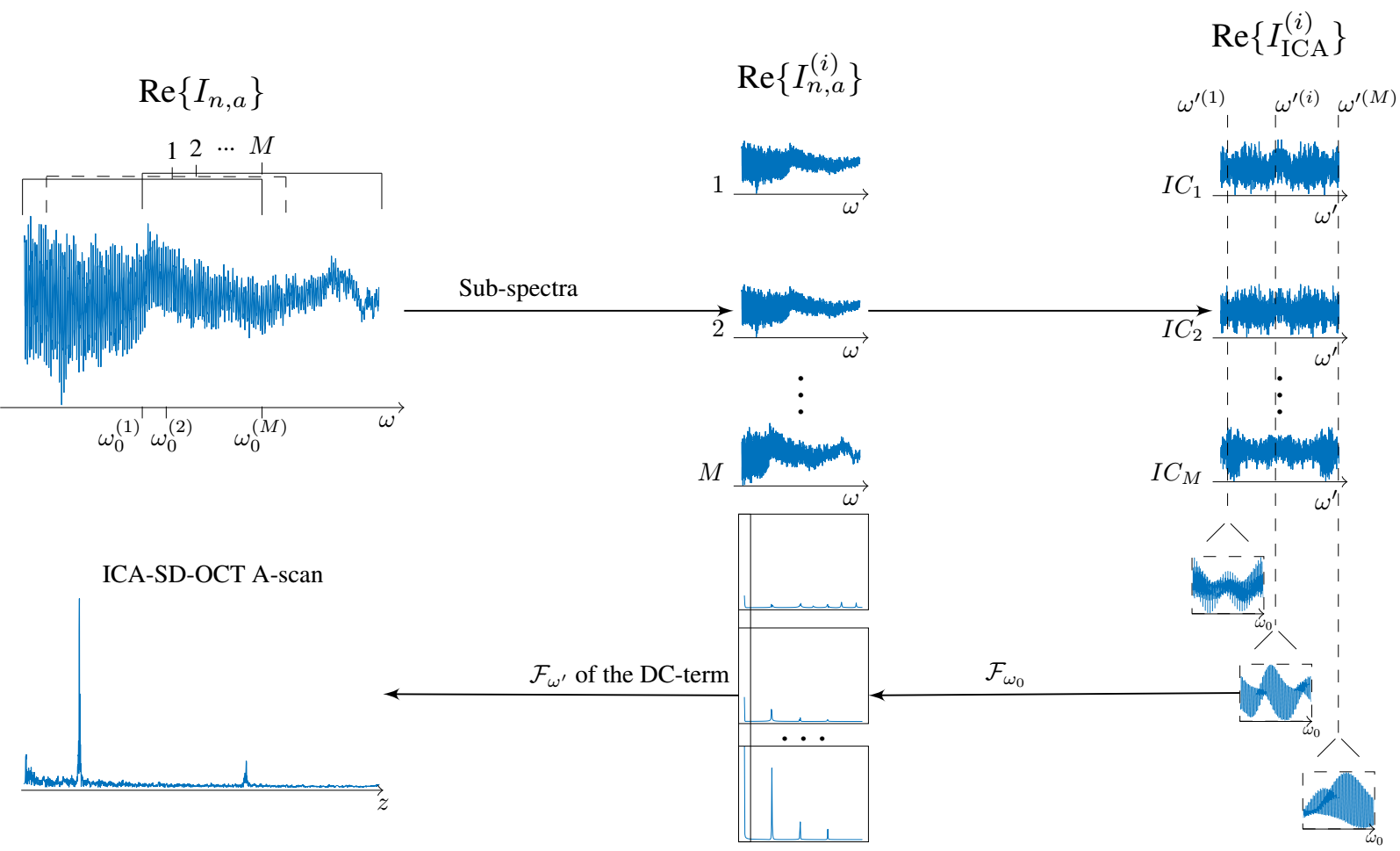}
\caption{Flowchart illustrating the artefact reduction algorithm. The analytic signal of the spectrum is divided into $M$ sub-spectra, which are treated individually according to equation \eqref{eq:ISI}. To obtain the final A-scan, the spectra are added and Fourier transformed in $\om$ (lower path). The upper path is equivalent to the lower path, but requires more computations.}
\label{fig:flowchart}
\end{figure}

The five last artefact terms in Eq. \eqref{eq:ISI} all have a $\sim \cos(\omega_0)$ dependence, either explicitly or implicitly through $\beta'$ because $\beta_0 = (\omega_0 n(\omega_0 ))/c$. To reduce the artefacts, we propose a procedure based on varying the centre frequency. This helps identify the artefacts \cite{banaszek2007blind,shirai2016improving}. Varying the centre frequency of the source is challenging, and therefore a numerical procedure is employed instead. A flowchart illustrating the process in seen in Fig. \ref{fig:flowchart} . The process works by numerically splitting the normalised, analytic spectrum $I_{n,a}$ of length $N$ into $M$ sub-spectra of length $N-M+1$, whose centres are shifted 1 pixel relative to their neighbours, as illustrated in panel 1 of Fig. \ref{fig:flowchart}. The first of the $M$ spectra comprises the first $N -M +1$ pixels of the full spectrum. The next sub-spectrum starts at pixel 2 of the full spectrum and so on, until sub-spectrum $M$, which is the last $N-M+1$ pixels of the full spectrum. This procedure varies numerically the centre frequency at the cost of narrowing the spectrum by $M-1$ pixels. This leads to a reduction of the resolution in $z$-space given that the width of the spectrum is limited by the spectrometer, which is the case for our source (see Methods). However, for $M\ll N$ the effect is limited. The ICA-SD-OCT procedure of Eq. \eqref{eq:ISI} is then applied to all $M$ sub-spectra independently, giving $M$ ICA-SD-OCT sub-spectra, shown in panel 2 of Fig. \ref{fig:flowchart}. These spectra correspond to a span of $\omega_0$’s with a fixed $\omega'$ axis. The $\omega_0$ span must be sufficiently large to ensure that artefact terms with $\sim \cos(\omega_0)$ dependence are removed when summing up the $M$ IC-SD-OCT sub-spectra. Applying a weighting function, $w(\omega_0)$ for this final summation, shown in panel 3 in Fig. \ref{fig:flowchart}, greatly reduces the $M$-value required for sufficient artefact reduction.
The weighting function weighs each sub-spectrum, such that the first has a lower weight than the second, and the central sub-spectrum has the highest weight. Because of the proportionality to $\cos(\omega_0)$ the artefacts oscillate in $\omega_0$, and without applying any weighting function, the limited span of $\omega_0$ causes the artefacts to exhibit severe side lobes in the $\omega_0$ Fourier domain, for which the Fourier transform is denoted $\mathcal{F}_{\omega _0}\lbrace I_{ICA}(\omega_0,\omega')\rbrace$. These side lobes can leak into the ICA-SD-OCT A-scan, which is found from the DC-term of  $\mathcal{F}_{\omega _0}\lbrace I_{ICA}(\omega_0,\omega')\rbrace$, and thus by reducing the side lobes of non DC-terms in  $\mathcal{F}_{\omega _0}\lbrace I_{ICA}(\omega_0,\omega')\rbrace$, the strength of the artefacts in the ICA-SD-OCT A-scan is likewise reduced. For optimal performance, the weighting function should exhibit weak or no side lobes, which can be achieved with e.g. a Gaussian or Hamming weight function. Formally, the artefact-free ICA-SD-OCT interferogram, $I_{ICA}^{(AF)}(\omega')$, is evaluated as 

\begin{equation}\label{eq:ICfinal}
I_{ICA}^{^(AF)}(\om) = \mathcal{F}_{\omega _0}\left\{ w(\omega_0)I_{ICA}^{(i)}(\om,\omega_0) \right\}(\om,0) = \int_{-\infty}^\infty w(\omega_0)I_{ICA}^{(i)}(\om,\omega_0) \mathrm{d}\omega_0.
\end{equation}
After the summing performed to implement Eq. \eqref{eq:ICfinal}, signals corresponding to the four terms in the second line of Eq. \eqref{eq:ISI} remain, and a Fourier transform with respect to $\omega'$ gives the even-order GVD-free ICA-SD-OCT A-scan, shown in panel 4 of Fig. \ref{fig:flowchart}. However, due to the spectral multiplication, all the reflection coefficients $r_1,r_2$ are also squared, and the resulting depth scan is thus a profile of the squared reflectivity instead of just the reflectivity. To re-obtain the OCT reflectivity profile (first order in reflectivity) the square root of the depth scans are evaluated.

\subsection*{OCT axial resolution in IC-OCT}
In the literature IC-OCT, both TD and SD, is generally claimed to have a $\sqrt{2}$ better axial resolution than conventional OCT \cite{kaltenbaek2008quantum,banaszek2007blind,lajunen2009resolution,ryczkowski2016experimental,shirai2014intensity,shirai2016improving,ogawa2016classical,Abouraddy2002QOCT}. However, we find this to be misleading because it originates from not defining the axial resolution from the same signal, i.e., conventional OCT defines it from the reflection profile, whereas IC-OCT defines it from the squared reflection profile. In IC-SD-OCT for example, the intensity spectrum is after the Michelson interferometer mirrored and combined with itself, whereas in for example chirped-pulse IC-TD-OCT two oppositely chirped pulses are combined. In other words, IC-OCT in general exploits fourth-order correlations in that it combines two intensity spectra, i.e., four complex field spectra, whereas conventional OCT exploits second-order correlation.  
However, if the original signal that is about to be squared in IC-OCT, cannot resolve two closely spaced reflectors but shows them as a single peak, then the squared signal will also only show a single peak. Thus, if the resolution was not defined as the FWHM of the A-scan of a single mirror, but as the distance between reflectors the system is able to resolve, then there would be no improvement in resolution with IC-OCT.
We would like to note that a “true” resolution improvement of $\sqrt{2}$ compared to standard (classical) OCT is found in so-called quantum OCT, which by nature requires two detectors and therefore inherently is IC-OCT, as demonstrated in \cite{nasr2003demonstration,Abouraddy2002QOCT}, after which a coincidence event is recorded varying the relative time delay (scanning the reference arm length similar to the procedure employed in TD-OCT), also known as the Hong-Ou-Mandel interferometer\cite{hong1987measurement}. The A-scan so obtained is in fact assimilated to a coincidence curve. For equal paths lengths a dip in the coincidence curve enabled by the unique temporal and spectral correlations between the two photons is observed. Due to the spectral entanglement between the two photons, the FWHM of this curve (assuming a Gaussian shape) is indeed a factor of $\sqrt{2}$ smaller than the FWHM of the A-scan of a mirror in conventional (classical) OCT, even when comparing the same orders of field reflectivity. This resolution improvement can only be explained by non-classical correlations between the two photons\cite{campos1990fourth,rarity1990two}.

%
%
%
%
\subsection*{Numerical simulations}
\begin{figure}
\centering

\includegraphics[width = 0.7\textwidth]{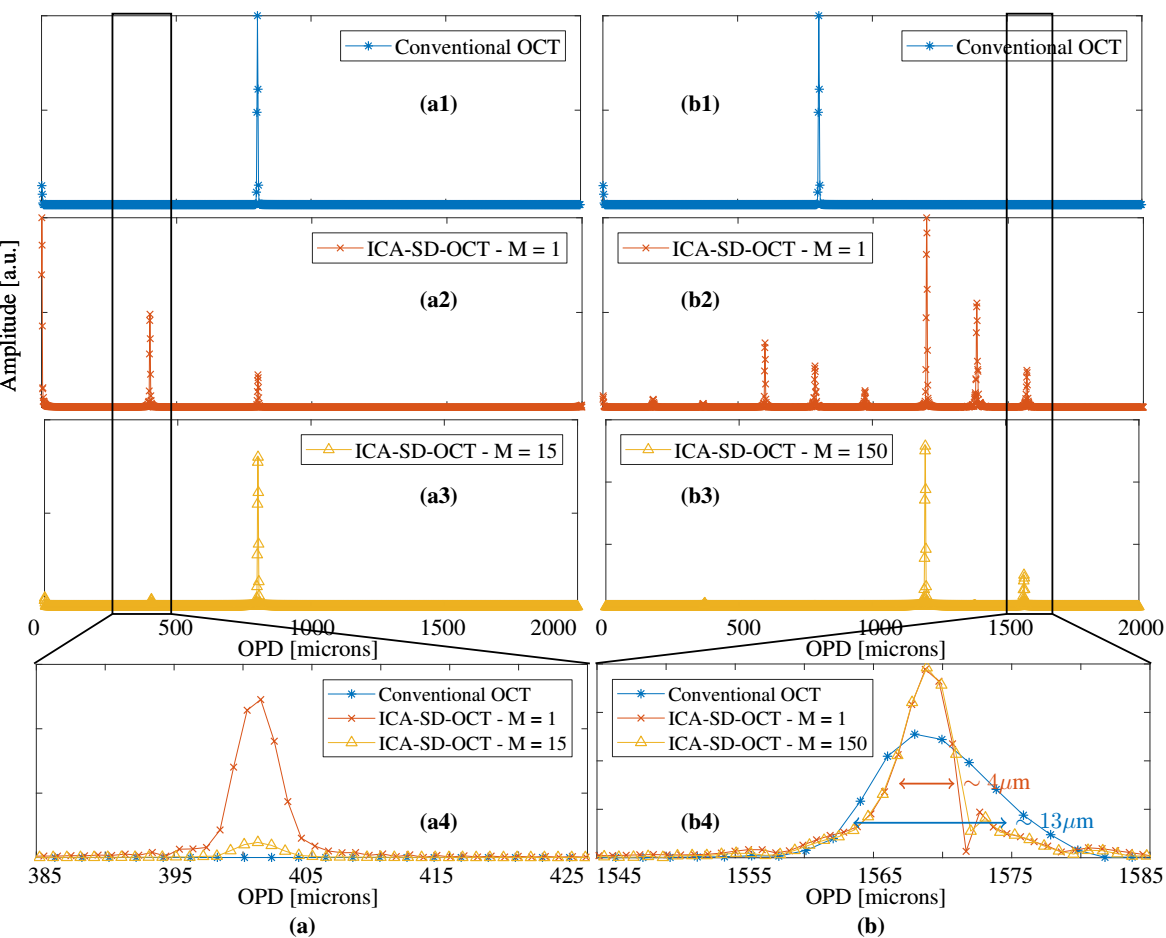}
\caption{Simulated data to illustrate the difference in results between conventional OCT and IC-OCT. \textbf{(a)} shows the case for a single reflector, and \textbf{(b)} shows the case for two reflectors. (x1)-(x3) shows conventional OCT, IC-OCT without the windowing procedure, and IC-OCT with the windowing procedure applied, respectively. (a4) and (b4) show zoom-ins of the artefact and the dispersion compensated peaks, respectively. The simulations are done with 2048 points spaced between 1070 nm and 1470 nm evenly in $k$-space, giving an imaging depth of 2 mm and a pixel distance of 1.97 microns. The source was simulated as a Gaussian spectrum with a central wavelength of 1300 nm and a FWHM of 230 nm.}
%
%
\label{fig:synth}
\end{figure}

%
%
We carried out a proof of principle simulation to test our ICA-SD-OCT approach. Figures \ref{fig:synth}(a1-a4) and (b1-b4) show the ICA-SD-OCT procedure applied to simulated data with one and two reflectors, respectively. For a single reflector conventional OCT is shown in Fig. \ref{fig:synth}(a1), ICA-SD-OCT without artefact reduction ($M=1$) is shown in Fig. \ref{fig:synth}(a2), and ICA-SD-OCT with $M=15$ artefact reduction is shown in Fig. \ref{fig:synth}(a3). An artefact emerging from the cross term between the single reflector and the DC-term is seen at $\sim 400$ microns in Fig. \ref{fig:synth}(a2), which is clearly suppressed by the $M=15$ artefact reduction, as seen in Fig. \ref{fig:synth}(a3) and the zoom in Fig. 3(a4). Figure 3(b) shows the result of a simulation of OCT imaging of a 100 micron thick silicon plate. The refractive index of silicon used for the simulation is the experimental data provided in \cite{li1980refractive} and then interpolated to fit our spectral pixels by a standard piecewise cubic Hermite interpolating polynomial (PCHIP) routine. Figures \ref{fig:synth}(b1)-(b3) shows the same as \ref{fig:synth}(a1)-(a3), but for two reflectors. In this case Fig. \ref{fig:synth}(b2) shows nine peaks including the DC-term (1 DC-term, 2 reflectors, 1 cross term from conventional OCT, and 5 ICA-SD-OCT artefacts), which corresponds directly to the nine terms in equation \eqref{eq:ISI}.
Figure \ref{fig:synth}(b1) constitutes the baseline for what is possible to achieve with ICA-SD-OCT in terms of artefact reduction. The two reflectors at 1200 microns and 1550 microns, stemming from the silicon plate, and the cross term at 350 microns are the three peaks that should be left after the IC-OCT windowing procedure. Figure \ref{fig:synth}(b3), which shows the result for ICA-SD-OCT with $M = 150$ artefact reduction, demonstrates that $M = 150$ is enough to suppress the ICA-SD-OCT artefacts and recover the 3 peaks from Fig. \ref{fig:synth}(b1). 



\section*{Methods}\label{sec:methods}
\begin{figure}
\centering
\includegraphics[width=1\textwidth]{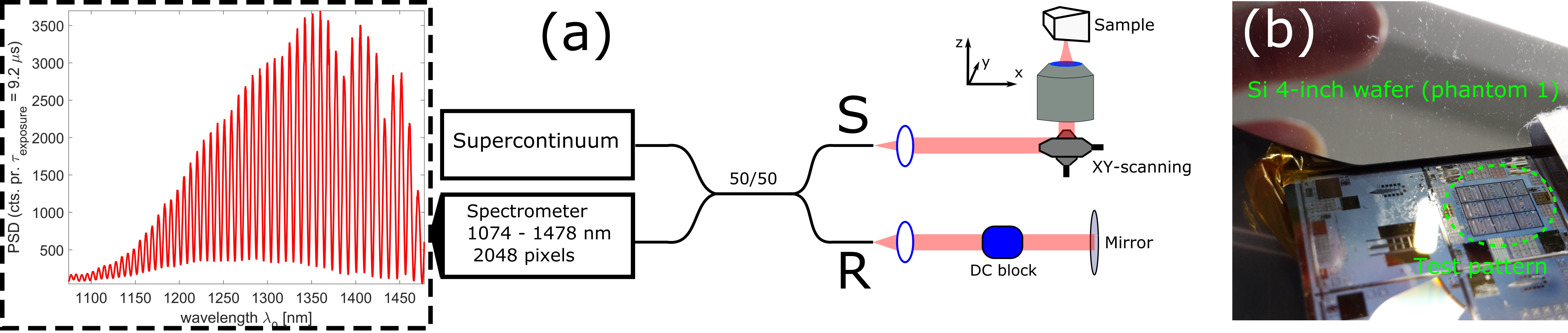}
\caption{(a) Simplified sketch of the experimental SD-OCT setup. The broadband NIR light is split evenly into a reference arm (R) and a sample arm (S). The interferometric signal is detected in the fourth arm of the coupler, and an example of an interferogram is displayed. (b) shows a photograph of one of the phantoms imaged.}
\label{fig:Setup}
\end{figure}

For imaging we used the conventional OCT system sketched in Fig. \ref{fig:Setup}. As optical source we used a 320 MHz superK Extreme EXR-9 OCT (NKT Photonics A/S, Denmark)  where a longpass filter selected light in the range $1000-1750$~nm. The high repetition rate supercontinuum is particularly suited for OCT \cite{maria17qswitch}. A 50/50 fibre coupler (Goosch and Housego, Netherlands) customized for the $1300$~nm wavelength band, served as the beam splitter and standard achromatic lenses collimated the light in each output arm. In the sample arm, galvanometer scanners are deployed for scanning of the sample through a microscope objective (LSM02, Thorlabs, UK). In the reference arm a block of glass was placed before the mirror for approximate hardware DC. Interferograms were recorded with a spectrometer C-1070-1470-GL2KL (Wasatch, USA) providing a $\sim 400$~nm bandwidth and operating at a line rate of 76 kHz. 

To eliminate the spectrometer non-linearity between wavenumber and pixel number, two reference interferograms are collected using a mirror as a sample, placed at two different axial positions as in \cite{makita2008full}. This technique can also be used for standard single-reflector DC, which we will compare with IC-SD-OCT all-depth multi-reflector DC in the following. With the standard DC applied, an axial resolution ranging from 3 microns to 5 microns (FWHM of Gaussian fit) over the 2~mm image range was measured (using a mirror as sample). Laterally we found our system to be able to distinguish features down to $6$~$\mu$m (USAF target 1951 phantom). For a power of $2.4$~mW on the sample, the sensitivity is $89$~dB. All interferograms are filtered with a Tukey window in $\omega'$ with bandwidth 300 nm centred at 1300 nm to smoothen the image. All A-scans and B-scans presented are single shot images with no temporal averaging applied.

\subsection*{Data availability}
The datasets generated during and/or analysed during the current study are available from the corresponding author on reasonable request.
\section*{Results}
To verify the theory and the results of the simulation, we imaged two phantoms. Standard DC, as described in the methods section, was applied only where mentioned explicitly. Phantom 1 is a polished silicon wafer of thickness 255 microns. The GVD of crystalline silicon is estimated to be $1100\pm100$ fs$^2$/mm \cite{li1980refractive}, which is sufficient to cause significant broadening of the interface corresponding to the bottom surface of the wafer. Assuming a Gaussian spectrum, the relative broadening factor, p, due to GVD is calculated as: 
\begin{equation}\label{eq:p}
p = \sqrt{1 + L^2\beta_2^2\left(\frac{\pi c \Delta \lambda}{\sqrt{2\ln(2)}\lambda_c^2}\right)^4}.
\end{equation}
Here $L$ is the physical axial position relative to the surface of the sample, in this case 255 microns. $\beta_2 = \partial^2 \beta /\partial \omega^2$ is the GVD parameter, $c$ is the speed of light in vacuum, and $\Delta \lambda$ and $\lambda_c$ are the full width at half maximum (FWHM) and centre wavelength, respectively. From the estimated GVD parameter we expect a relative broadening of the bottom surface by a factor of $\sim 4.1 \pm 0.3$. 

\begin{figure}
\centering
\includegraphics[width = 0.8\textwidth]{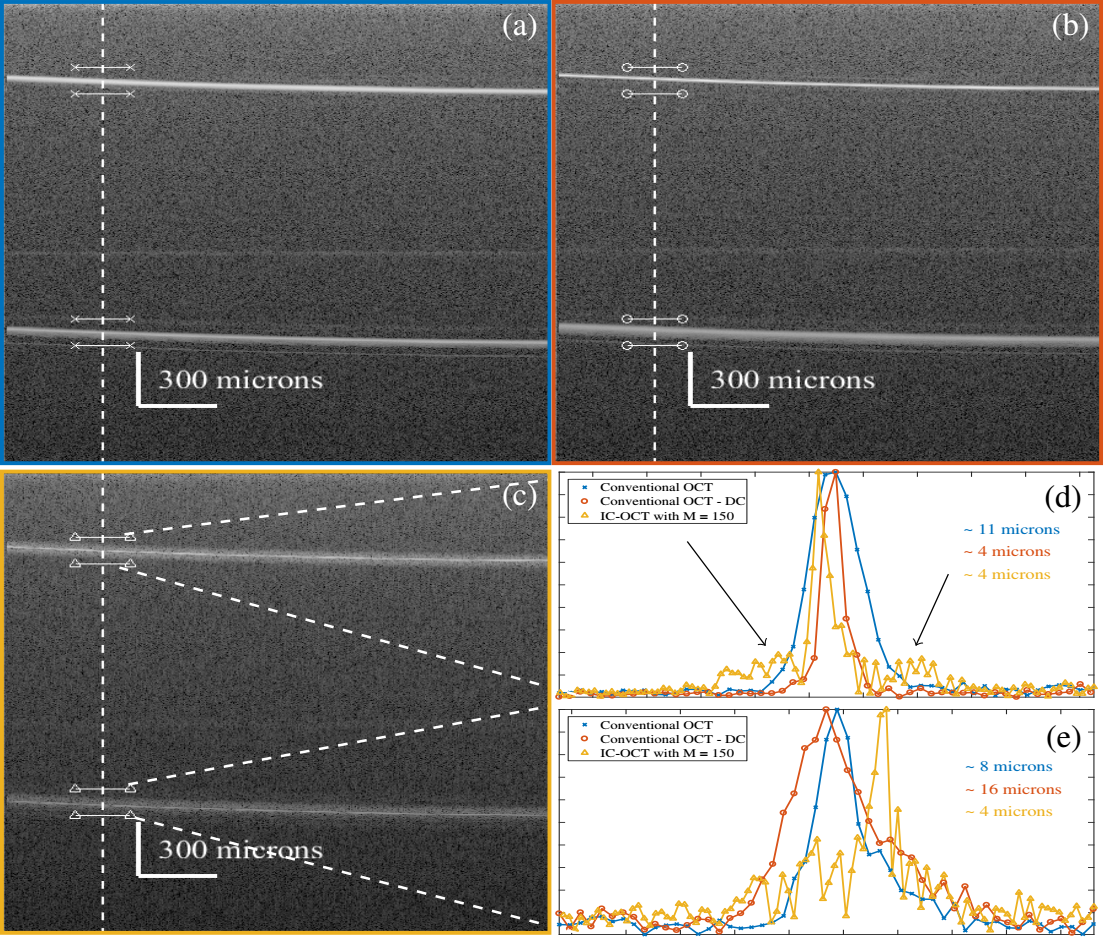}
\caption{B scan of a polished silicon wafer with (a) no dispersion compensation, (b) global compensation of system dispersion, and (c) IC-OCT image with $M=150$. (d) shows pieces of the A scan along the white dashed line in (a)-(c). All images are single shot and filtered with a Tukey window.}
\label{fig:Bwafer}
\end{figure}

Cross sectional images, B-scans, of the silicon wafer are shown in Fig. \ref{fig:Bwafer}, with \ref{fig:Bwafer}(a) being the image collected without any dispersion compensation, \ref{fig:Bwafer}(b) the image with standard DC of the top interface, and \ref{fig:Bwafer}(c) the ICA-SD-OCT image. Figures \ref{fig:Bwafer}(d) and \ref{fig:Bwafer}(e) shows the A-scan profile along the vertical dashed lines between the short horizontal, solid lines for the top and bottom surface of the wafer, respectively. The image with no DC in Fig. \ref{fig:Bwafer}(a) shows the two surfaces having approximately the same thickness of  $\sim 10$ microns despite the highly dispersive sample. The top interface is broadened due to the dispersion in the set-up, while the bottom interface is broadened by the combined effect of the dispersion in the set-up  and in the sample. As the set-up dispersion and sample dispersion have different signs, the accumulated dispersion for the bottom interface is in magnitude smaller than the set-up dispersion, and therefore the bottom interface is thinner in the image than the top interface (but one is not always that lucky!). The image in Fig. \ref{fig:Bwafer}(b) displays a narrow top interface with a FWHM of 4 microns and a bottom interface with a thickness that has increased by a factor of approximately 4 to 16 microns, as expected from Eq. \eqref{eq:p}. The extra broadening of the bottom surface is due to the set-up dispersion having been cancelled, and it highlights the major drawback of conventional dispersion compensation: All depths cannot simultaneously achieve the theoretical dispersion-free axial resolution. As shown in Fig. \ref{fig:Bwafer}(c) the ICA-SD-OCT method allows thinning of all interfaces to about 4 microns simultaneously, irrespective of depth, by intrinsic cancellation of all even order GVD terms. The ICA-SD-OCT image is created with $M = 150$ sub-spectra, which allows to obtain significant reduction of the artefacts originating from cross terms between two scatterers seen in Figs. \ref{fig:synth}(a2) and \ref{fig:synth}(b2), with no trace of these artefacts even on a logarithmic scale. We note however that the ICA-SD-OCT procedure introduces a weak set of artefacts in the background of each reflector peak, seen as the blur around both surfaces in Fig. \ref{fig:Bwafer}(c), and as shoulders in Fig. \ref{fig:Bwafer}(d), as marked by the black arrows. These artefacts come from the cross term between a scatterer and the background noise, which was not included in the theoretical derivation or the numerical simulations. 

\begin{figure}
\centering
\includegraphics[width = 0.8\textwidth]{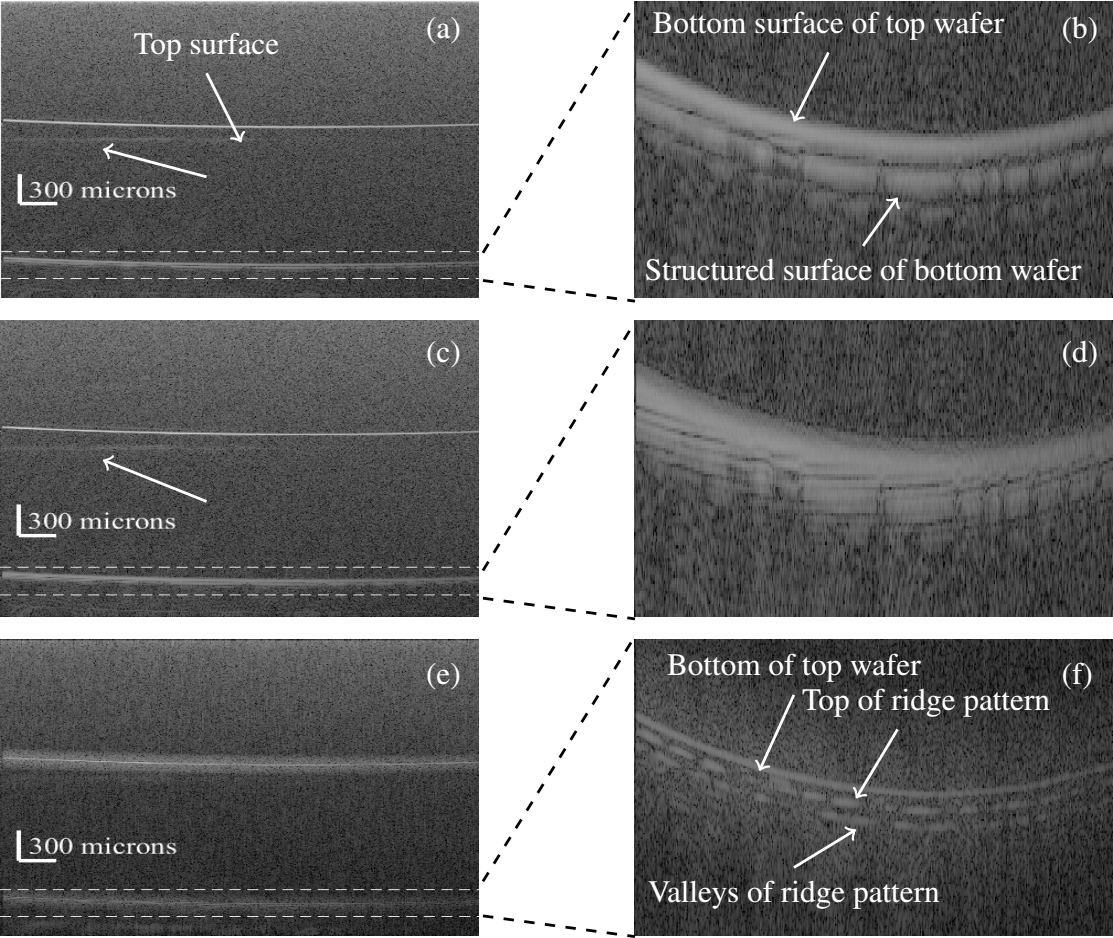}
\caption{B scan of a silicon wafer with surface structure placed beneath an ordinary silicon wafer. (a), (c), (e) shows images with no DC, global DC, and IC-OCT with $M=50$, respectively. (b), (d), and (f) shows zoom-ins of the bottom interface. The surface structure is clearly better resolved with IC-OCT.}
\label{fig:BwaferPattern}
\end{figure}

To further evaluate the performance of the ICA-SD-OCT procedure, we created phantom 2 by placing a silicon wafer with a surface structure directly below phantom 1. The B-scans are seen in Fig. \ref{fig:BwaferPattern} with (a), (c), and (e) being the image with no dispersion compensation, the image with compensation of set-up dispersion, and the ICA-SD-OCT image, respectively. \ref{fig:BwaferPattern}(b), (d), and (f) are zoom-ins of the interface between the top wafer and the structured surface wafer on the bottom.

This phantom imitates a sample with several interfaces and small scale features deep inside it. From Fig. \ref{fig:BwaferPattern}(a) and \ref{fig:BwaferPattern}(b) we see how a conventional OCT system with no dispersion compensation is not able to image the small scale features on the bottom wafer. In Fig. \ref{fig:BwaferPattern}(c) and (d)  we see that conventional dispersion compensation does not improve the resolution of the back surface (as in Fig. \ref{fig:Bwafer}) and even blurs the small features below the back surface even more than when no DC is applied. Both without dispersion compensation and with conventional single-reflector DC the the dipersion remains too high to image the structure on the bottom wafer.

In contrast, in Fig. 6(e)-(f), the ICA-SD-OCT image with $M = 50$ artefact reduction provides such a good all-depth dispersion compensation that the top wafer and the bottom wafer are clearly distinguished, allowing to see both the tops and valleys of the bottom structure. Clear imaging of the fine structures in Fig. \ref{fig:BwaferPattern}(f) demands sufficiently high axial resolution, and this requires a sufficiently high bandwidth, i.e. bandwidth of the $M$ sub-spectra. Consequently, $M$ cannot be large, which is why $M = 50$ was used. The relatively low $M$ value does not allow full elimination of the artefacts around each interface.
The well-known artefact just below the top surface in \ref{fig:BwaferPattern}(a) and \ref{fig:BwaferPattern}(c), as marked by arrows, originates from the autocorrelation term of the two main reflectors (term 3 in Eq. \eqref{eq:convOCT}), and therefore they cannot be reduced by the protocol presented. However, the artefacts are reduced by the IC-OCT procedure itself, and they are not visible in Fig. \ref{fig:BwaferPattern}(e). In the case of a more complex sample with several scattering centres and intertwining structures, the $M$ value must be chosen large enough such that the signal background noise is sufficiently reduced.

\section*{Summary and Conclusions}
In summary, we have theoretically and experimentally demonstrated a new ICA-SD-OCT procedure that allows all-depth dispersion compensation of all even order GVD. We show that we numerically can eliminate the GVD due to the sample, irrespective of the scattering depth, allowing us to maintain the theoretical axial resolution at all depths, and this using a conventional SD-OCT set-up with only a single spectrometer. Furthermore, our new numerical procedure maintains the axial range as well as generically removes all artefacts emerging from the multiplication of two interferograms, which was not possible before. Numerical simulations were performed using a single and dual layer sample to investigate the dispersion compensating abilities and artefact reduction of the ICA-SD-OCT procedure. These simulations demonstrated tolerance to GVD from the sample as well as excellent reduction of the artefacts. Two phantoms were imaged experimentally, a single polished silicon wafer, and the same polished silicon wafer with another structured silicon wafer placed underneath it. We demonstrated how a conventional SD-OCT system with conventional single-reflector dispersion compensation, showed a severely broadened bottom surface due to sample dispersion and was not able to image the surface structure of the bottom wafer. In contrast our experimental results clearly showed how an ICA-SD-OCT system could compensate dispersion in-depth and image the bottom small features 260 microns into the phantom and re-establish a 4 micron resolution for both top and bottom surfaces. 
%
%
%
%

\bibliography{References.bib}







\end{document}